\documentclass[twoside,letter]{elsart} 
\journal{Astroparticle Physics}
\usepackage[dvips]{graphicx}
\usepackage[dvips]{color}
\usepackage{natbib}
\usepackage{amssymb}  
\usepackage{wasysym}  
\definecolor{light}{gray}{0.98}
\definecolor{Light}{gray}{0.93}
\definecolor{Lightblue}{rgb}{0.2,0.2,0.40}
\usepackage[latin1]{inputenc}                
\usepackage[T1]{fontenc}                     
\begin{document}
\begin{frontmatter}
\title{The diffuse neutrino flux from FR-II radio galaxies and blazars:\\ A source property based estimate}
\author[dort]{Julia K.~Becker\corauthref{cor}} 
\author[bonn,bonn2]{, Peter L.~Biermann}
\author[dort]{, Wolfgang Rhode}
\corauth[cor]{{\scriptsize Corresponding author. Contact: julia@physik.uni-dortmund.de, phone: +49-231-7553667}}
\address[dort]{Department of Physics, Dortmund University, D-44221 Dortmund, Germany}
\address[bonn]{Max Planck Institut f\"ur Radioastronomie, Auf dem H\"ugel 69,
  D-53121 Bonn, Germany}
\address[bonn2]{Department of Physics and Astronomy, University of Bonn, Germany}
\date{\today}
\begin{abstract}
Water and ice Cherenkov telescopes of the present and future
  aim for the detection of a neutrino signal from
  extraterrestrial sources at energies $E_{\nu}>$PeV [\cite{AMANDA,Antares,icecube}]. Some of the most promising extragalactic
  sources are Active Galactic Nuclei (AGN).
 In this paper, the neutrino flux from
  two kinds of AGN sources will be estimated assuming $p\,\gamma$
  interactions in the jets of the AGN. The first analyzed sample
  contains FR-II radio galaxies while the second AGN type examined are
  blazars. The result is highly dependent on the proton's index of the
  energy spectrum. To normalize the spectrum, the connection between neutrino
  and disk luminosity will be used by applying the jet-disk symbiosis model
  from~\cite{falcke1}. The maximum proton energy and
  thus, also the maximum neutrino energy of
  the source is connected to its disk luminosity, which was shown
  by~\cite{lovelace} and was confirmed by~\cite{falcke}.\\
\end{abstract}
\end{frontmatter}
\section{Introduction}
Large volume neutrino telescopes like AMANDA, Antares and IceCube [\cite{AMANDA,Antares,icecube}] are being built to detect the 
neutrino flux from extraterrestrial sources. For a diffuse analysis, the detection of such a signal is possible at energies at which the extragalactic neutrino
flux contribution exceeds the number of atmospheric neutrinos. Active Galactic Nuclei (AGN) are believed to make up a large fraction
of the extragalactic neutrino flux. While previous models - see e.g.~[\cite{learned}] for a review - used the Cosmic Ray flux or a certain part of the photon spectrum 
to estimate the neutrino flux from AGN, the model presented in this paper uses the jet-disk symbiosis
model by~\cite{falcke1} and allows for the dependence of the maximum particle
energy on the source properties. This is the key point of our analysis.

The total
neutrino spectrum at Earth is given as
\begin{equation}
\Phi(E_\nu^0)=\int_{z}\,\int_{L} dz \, dL\, \frac{dN}{dE_{\nu}}(E_{\nu}^0,L,z)\cdot \frac{dn}{dL}(L,z) \cdot \frac{dV}{dz}\cdot \frac{1}{4\pi d_L(z)^2}\,.
\label{flux_equation}
\end{equation}
The parameters in the calculation are the integral radio luminosity $L$ and the
redshift $z$. The energy of the neutrinos at the source $E_{\nu}$ is
redshifted due to adiabatic energy losses on the particles' way
toward Earth, $E_{\nu}=(1+z)\cdot E_{\nu}^{0}$, where $E_{\nu}^{0}$
is the neutrinos' energy at Earth. $dN/dE_{\nu}$ is the single source
spectrum of an AGN, while $dn/dL$ is the radio luminosity function
(RLF) per comoving volume $dV/dz$ and luminosity interval. To obtain
the total number of AGN per redshift interval, the RLF has to be
weighted by the comoving volume and divided by a factor of
$4\,\pi\,d_L^2$ with $d_L$ as the luminosity distance to receive the neutrino
flux at Earth. Finally,
the result has to be integrated over all possible redshifts and
luminosities, where the integration limits depend on the source
sample as discussed in section~\ref{limits}.\\
The neutrino flux will be calculated for two different source samples,
one including FR-II sources with strong extended radio emission (in the following referred to as ''steep spectrum sources'')
and the second one flat spectrum blazar sources (referred to as ''flat spectrum sources'' throughout the paper). It will be assumed that
neutrinos are produced by pions resulting from a $\Delta$-resonance of
the initial $p\,\gamma$ interaction. In this scenario, the neutrino
spectrum follows the proton spectrum. Multipion events (see e.g.~[\cite{muecke99}]) will not be
  included in this calculation, since the changes of the spectrum due to these
  events would lie within the estimated uncertainty of the calculation due to uncertainties in the
  jet-disk symbiosis model and the sources' Radio Luminosity Functions (RLFs), see section~\ref{uncertainties}.
For the steep spectrum sources, the spectral index of the proton
spectrum will be calculated from the synchrotron emission of the
electrons. This is possible since the proton and electron spectra
resulting from
shock acceleration are equivalent. This method cannot be used for the
flat spectrum source sample, since the radio emission of the jet is
spatially unresolved and therefore, the observed spectrum is a
superposition of the different radio spectra. In this case, the spectrum will be a superposition of the spectra of all knots, each with a particle
index of $p\simeq 2$, following the derivation of the index in
diffuse shock acceleration, see e.g.~[\cite{prothfermiacc}]. 
The maximum energy of the protons in the jets will
be assumed to be luminosity dependent as is suggested in~[\cite{lovelace}]. To normalize the generic AGN energy spectrum, the
jet-disk symbiosis model given in~[\cite{falcke}] will be
applied to connect the neutrino luminosity with the disk
luminosity and therefore, also with the radio luminosity of
the jets. Throughout this paper, an Einstein-de Sitter Universe ($\Omega_m=1,\,\Omega_\Lambda=0$) with
$h=0.5$ is assumed, since the used RLFs are not available in the 
concordance model ($\Omega_m\approx 0.3,\,\Omega_{\Lambda}\approx 0.7$) which
has been confirmed by recent measurements [\cite{WMAP}]. However, the result should be independent of the
cosmology used as it has been pointed out in~[\citet{dunlop}].\\
The paper is organized as follows: 
In section~\ref{samples}, the two AGN samples will be discussed. They
will be
used to determine the AGN radio
luminosity function (RLF). In case of the steep spectrum sources, the spectral
indices between $2.7$ and $5$~GHz are
used to evaluate the index of the primary particle spectrum. 
Section~\ref{single_source} gives an explicit description of
the normalization of the single source spectrum to the disk luminosity
of the objects. 
In section~\ref{limits}, the integration limits for the redshift and
luminosity distribution are discussed. Finally, the different models for both types
of AGN sources are applied to calculate the AGN neutrino flux. It will
be shown that the result strongly depends on the index of the particle spectrum.
\section{The samples \label{samples}}
The first sample consists of 356 steep spectrum sources, selected by~\cite{willott}. The sources are sampled from the
7CRS [\cite{McGil,lacy}], 6CE [\cite{rawlings}] and 3CRR [\cite{laing}]
catalogs with flux measurements at $178$~MHz. 
The second sample  comprises $171$ flat spectrum sources from~\cite{dunlop}. 
They are selected from the Parkes selected regions sample, the Parkes $\pm 4^{\circ}$ zone [\cite{peacock,wright}], the 'Northern-Sky' survey of~\cite{PeacockWall} and the 'All-Sky' survey 
of~\cite{WallPeacock}. 
\subsection{The spectral and particle index of the sources}
As discussed in detail in, e.g.,~[\cite{rybicki}], the
index $p$ of the proton spectrum is correlated to the index $\alpha$ of synchrotron
radiation by $p = 2\alpha+1$.
\subsubsection{Steep spectrum sources}
\begin{figure}[h]
\centering{
\includegraphics[width=10cm]{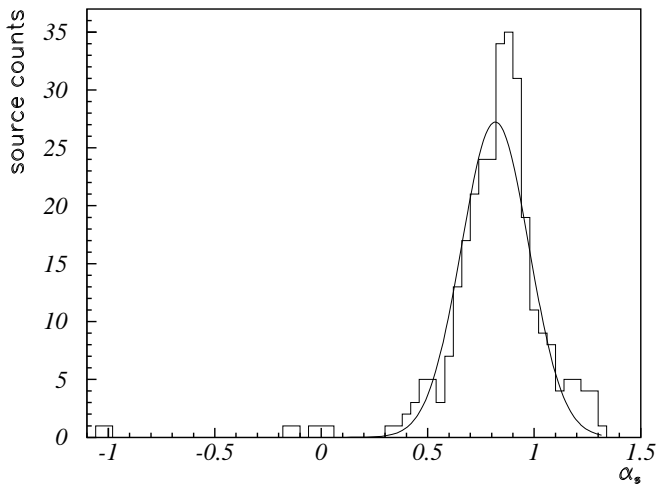}
}
\vspace{-0.5cm}
\caption[Histogram of the spectral indices of the steep spectrum sources.]
{Histogram of the spectral indices of the steep spectrum sources. Sources counts are shown against the spectral index at 
$5$~GHz. The maximum is found at $\alpha_s=0.8$. The width of the Gau{\ss} distribution is $\sigma_s=0.2$.}
\label{synindex_willott}
\end{figure}

123 of the total 356
sources are identified in the S1-S5 catalogs [\cite{s1s2,s3,s4,s5}] which give the spectral index of the sources 
at $2.7-5$~GHz.
To estimate the steep spectrum index $\alpha_s$ at $2.7-5$ GHz, the maximum in a distribution of the source counts versus $\alpha_s$ is determined applying a Gau{\ss} fit (see Fig.~\ref{synindex_willott}). 
The maximum is found to be at $\alpha_s=0.8$
with a peak width of $\sigma_s=0.2$. In calculations of the AGN evolution functions by Willott et al., a spectral index of $\alpha_s=0.8$ is assumed~[\cite{willott}]. 
The spot check of the sources supports this assumption. Hence, the index of $\alpha=0.8$ is adopted. 
This result is supported by~\cite{gregorini} who analyze the spectral index of radio galaxies 
in the range of $2.7$~to~$5$~GHz resulting in an average index of around $0.8$ for steep spectra.
Consequently the particle index is 
$p_s=2\cdot \alpha_s+1=2.6$. Here, it is assumed that the stationary
  particle spectrum is equal to the particle injection spectrum, which is
  based on the assumption that particles from this region undergo energy scale
  independent losses such as would occur in a flow. There is unfortunately no
  good and well-tested model available at this  time for the
particle transport, convection versus diffusion and other processes, so
that we make this simple assumption by Occam's razor.
To demonstrate the influence of the particle index on the resulting 
spectrum, the calculation will also be done with different harder
spectral indices.
\subsubsection{Flat spectrum sources}
Since the radio spectrum of blazars is spatially unresolved, the
observed index of the radio spectrum cannot be identified with the
index of synchrotron radiation. The particle spectrum is a superposition of the spectra of the single knots,
\[
dN/dE_{\nu}\propto \sum_{i} a_i\, E^{-p}
\]
assuming the same particle index for all sources, $p\simeq2$, as suggested by
diffuse shock acceleration for strong shocks, and also by the radio data of strong jets. 
An index close to $p\simeq2$ is supported by the individual radio spectra as described in, e.g., [\cite{bridle}] for strong sources. 
\subsection{Radio Luminosity Functions}
\subsubsection{Steep spectrum sources}
\begin{figure}
\setlength{\unitlength}{1cm}
\begin{minipage}[t][9.5cm][b]{6.7cm}
\begin{picture}(6.7,3.5)
\includegraphics[width=7.5cm]{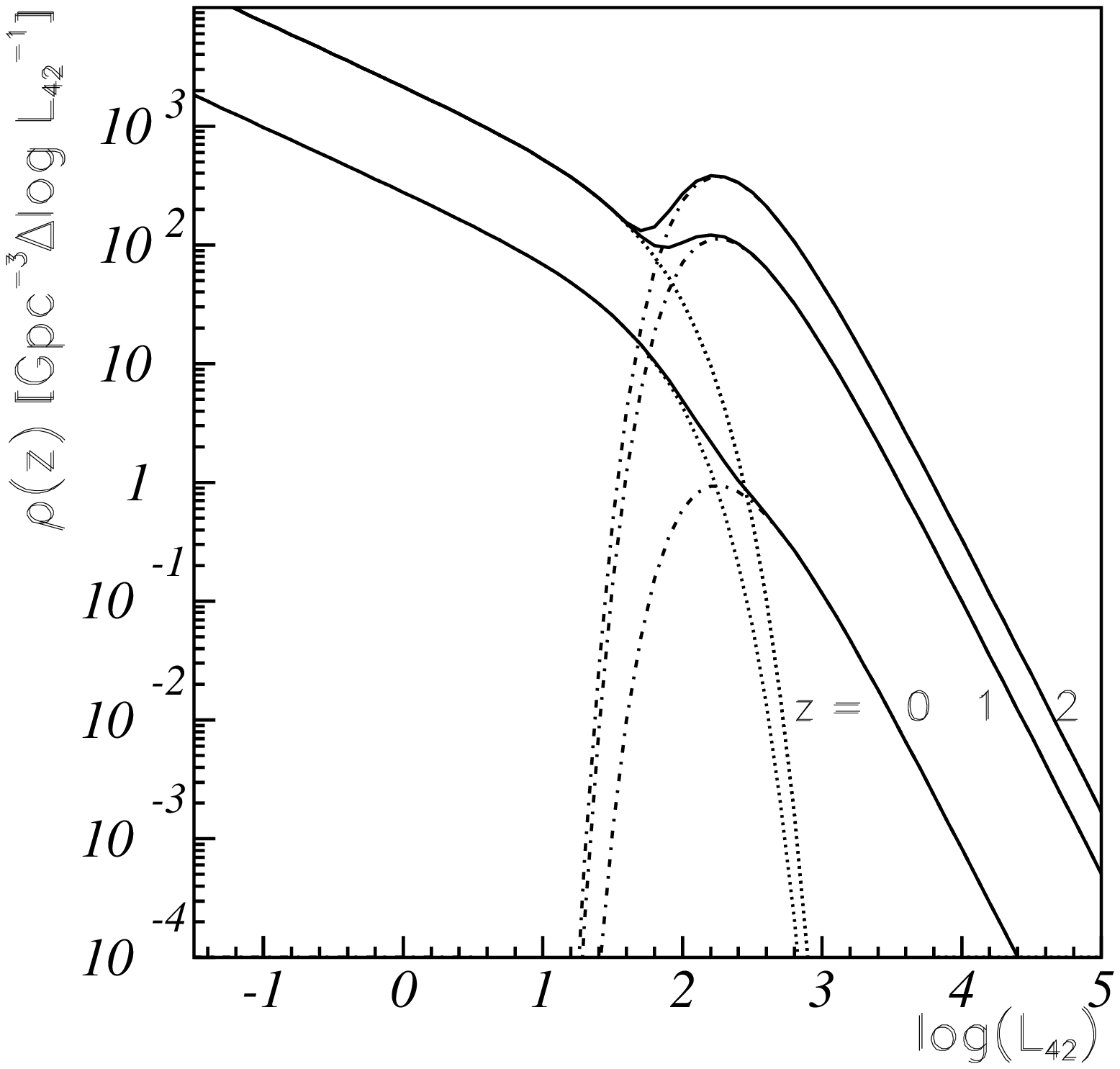}
\end{picture}\par
\caption[]{Steep spectrum RLF according to~\cite{willott} versus luminosity for $z=0,\,1,\,2$.
\\[0.5cm]}
\label{willott_lum_abst_paper}
\end{minipage}
\parbox{0.5cm}{\quad}
\begin{minipage}[t][9.5cm][b]{6.7cm}
\begin{picture}(6.7,3.5)
\includegraphics[width=7.5cm]{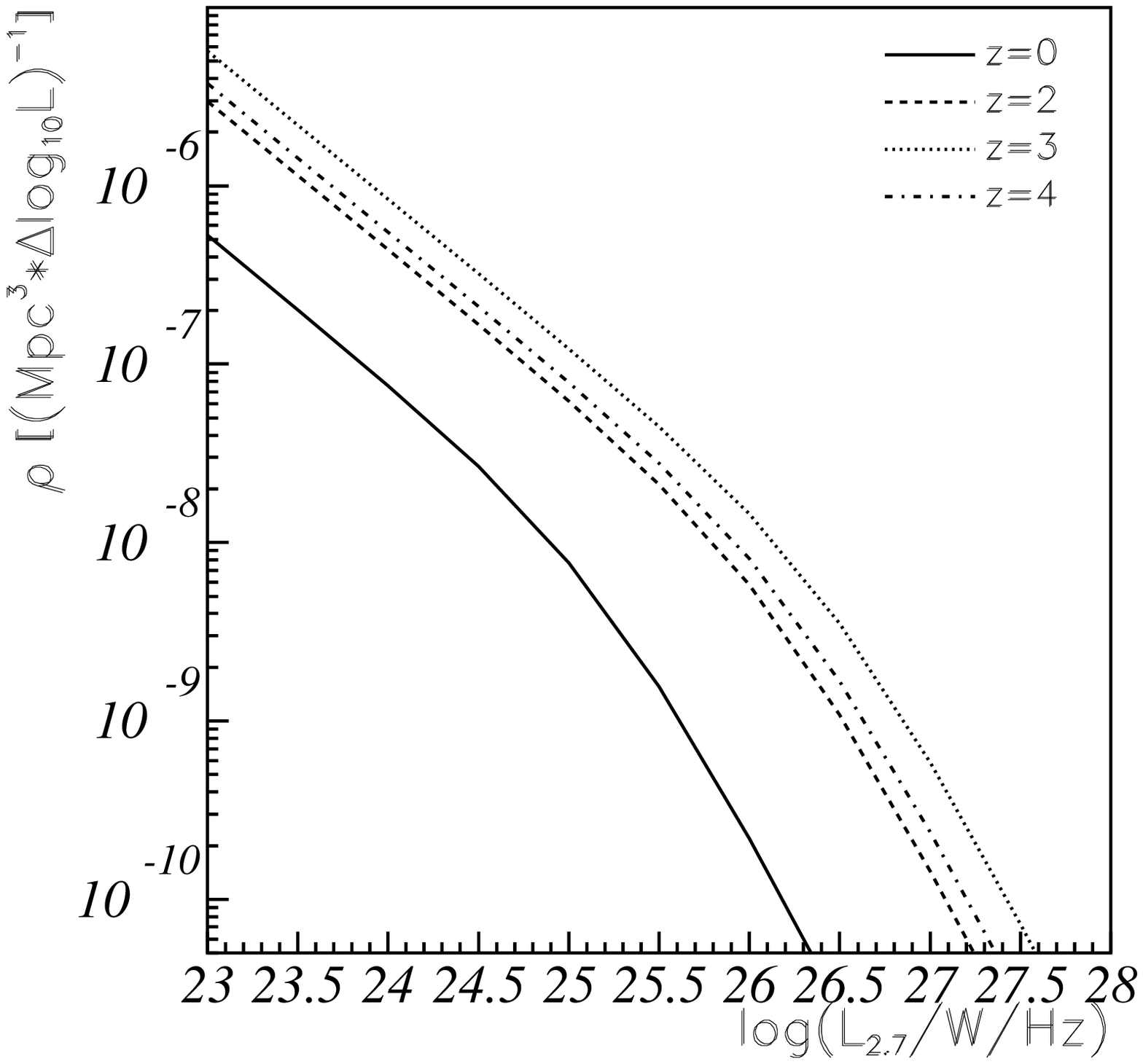}
\end{picture}\par
\caption{RLF for flat spectrum sources in a pure luminosity evolution model according to~\cite{dunlop}. Four different redshift values are used, $z=0,\,2,\,3,\,4$.}
\label{peacock_rlf_paper}
\end{minipage}\hfill
\end{figure}

The AGN RLF depends on the luminosity and on the redshift. 
To find the proper relation empirically, a factorizing separation of
the density in a luminosity dependent function and a redshift
dependent distribution is assumed by~\cite{willott}. The model includes the steep spectrum sources as they are explained above. 
The AGN RLF is given at a frequency of
$0.151$~GHz depending on the differential luminosity $L_{0.151}$.
The RLF was assumed to consist of two separate distributions,
a low luminosity function including objects without or with only weak emission lines and
a high luminosity function with objects with strong emission lines. 
The density $\rho(L_{0.151},z)$ is given as the product of a pure luminosity function $\rho_l(L_{0.151})$ in units of 1/(Gpc$^3$\,$\Delta \log L_{0.151}$) 
and the dimensionless evolution function $f(z)$:
\[
\rho(L_{0.151},z)=\rho(L_{0.151})\cdot f(z)\,.
\]
Two AGN populations contribute to the RLF:\\
\emph{Low luminosity sources} only show weak or no emission lines. These sources are
 FR-I and weak emission line FR-II galaxies. 
Above a luminosity $L_{0.151}\approx 10^{25.5}$~W/(Hz~sr), the sources can almost exclusively be 
assumed to be FR-II galaxies. For the low luminosity function, $\rho_l(L)$, an ansatz of
\[
\rho_l(L_{0.151})=\rho_{l}^{0}\left( \frac{L_{0.151}}{L_{l}^{*}}\right)^{-\alpha_l}\cdot \exp\left[ -\frac{L_{0.151}}{L_{l}^{*}}\right]
\]
is made. This analytic ansatz is known as the Schechter function in literature [\cite{schechter}].
$\alpha_l$ is the power law shape, $\rho_{l}^{0}$ is the normalization constant and $L_{l}^{*}$ is the break luminosity. The evolution function of the low luminosity population is taken to be
\[
f_l(z)=\left\{ 
\begin{array}{ccc}
(1+z)^{k_l}&&\mbox{ for } z \leq z_{l}^{0} \\
(1+z_{l}^{0})^{k_l}&&\mbox{ for } z >  z_{l}^{0} \,.
\end{array}
\right.
\]
The redshift evolution of the sources is known up to $\sim z_{l}^{0}$. At higher redshifts, it is assumed
to be constant since there
are experimental indications [\cite{miyaji}] of a constant or slightly
decreasing distribution at redshifts up to $z~\sim 6$. 
The values of five free parameters used in the functions are listed in table~\ref{willott_params}. 
The complete (luminosity and redshift dependent) RLF is shown in Fig.~\ref{willott_lum_abst_paper}.\\
\emph{The high luminosity population} consists of FR-II galaxies with strong emission 
lines. The form of the high luminosity function, $\rho_h(L_{0.151})$, 
is similar to the one for the low luminosity population. Here, however, the exponential function applies
at low luminosities while the power law dominates at higher luminosities:
\[
\rho_{h}(L_{0.151})=\rho_{h}^{0}\left( \frac{L_{0.151}}{L_{h}^{*}} \right)^{-\alpha_h}\cdot \exp\left[-\frac{L_{h}^{*}}{L_{0.151}}\right]\,.
\]
The evolution function of the high luminosity population is assumed to be an exponential function up to a certain redshift $z_{h}^{0}$ and then continuing as a constant:
\[
f_{h}(z)=\left\{ 
\begin{array}{ccc}
\exp\left[-\frac{1}{2}\left( \frac{z-z_{h}^{0}}{z_{h}^{1}}\right)^2 \right]&&\mbox{ for } z\leq z_{h}^{0}\\
1&&\mbox{ for } z> z_{h}^{0}\,.
\end{array}
\right.
\]
A 
comparison to X-ray data~[\cite{miyaji}] justifies an exponentially increasing 
evolution. 

The luminosity, which is here given at $0.151$~GHz, is used in the integral form for further calculations. Thus the luminosity $L_{0.151}$ is converted to the integral radio luminosity in units of $10^{42}$ erg/s, $L_{42}$, by assuming a spectral index of $\alpha=0.8$. 
Furthermore, the RLF is given in units of Gpc$^{-3}\,(\Delta
\log(L))^{-1}$. 
In following calculations, the RLF will be given per luminosity
interval $dL_{42}$. 
\begin{table}[h]
\centering{
\begin{tabular}{|c|c|c|c|c|c|c|c|c|c|}
\hline
$\rho_{l}^{0}$&$\alpha_l$&$L_{l}^{*}$&$z_{l}^{0}$&$k_l$&$\rho_{h}^{0}$&$\alpha_h$&$L_{h}^{*}$&$z_{h}^{0}$&$z_{h}^{1}$\\ \hline\hline
$10^{1.850}$&$0.542$&$10^{26.14}$&$0.646$&$4.10$&$10^{-6.260}$&$2.31$&$10^{26.98}$&$1.81$&$0.523$\\ \hline
\end{tabular}
}
\caption{Parameters used in the calculation of the AGN RLF for steep spectrum sources [\cite{willott}]. 
$\rho_{l}^{0}$ and $\rho_{h}^{0}$ are given in units Gpc$^{-3}\Delta(\log L)$. $L_{h}^{*}$ and $L_{l}^{*}$ are given in~W/(Hz sr).}
\label{willott_params}
\end{table}

\subsubsection{Flat spectrum sources}
$171$ flat sources from four catalogs at $2.7$~GHz are used by~\cite{dunlop} to determine the RLF of flat spectrum
blazars as described above. The ansatz for the
flat spectrum RLF is a pure luminosity evolution function of the form
\[
\rho(L,z)=\rho_0\cdot \left\{\left( \frac{L}{L_c(z)}\right)^{\alpha}+\left( \frac{L}{L_c(z)}\right)^{\beta} \right\}^{-1}\,.
\]
The
function is labeled pure luminosity evolution since only the break
luminosity $L_c$ evolves with the redshift.
Here, $\alpha$ and $\beta$ are the power law slopes, $\rho_0$ is the normalization and $L_c(z)$ is the break luminosity which evolves with redshift:
\[
\log[L_c(z)]={a_0+a_1\cdot z+a_2\cdot z^2}\,.
\]
The luminosity $L$ is given in units of W/(Hz sr) at a frequency of $2.7$~GHz. The RLF is
shown in Fig.~\ref{peacock_rlf_paper} with the set of parameters as given in table~\ref{peacock_params}. For further calculations, the
luminosity at $2.7$~GHz is converted to an integral luminosity by
assuming a flat spectrum with $\alpha_f=0$. The RLF is given in units of Gpc$^{-3}\cdot(\Delta\log L)^{-1}$, and thus the differential RLF is
\begin{equation}
\frac{dn}{dL}(L,z)=\frac{\rho_0}{\ln(10)\cdot L}\cdot \left\{\left( \frac{L}{L_c(z)}\right)^{\alpha}+\left( \frac{L}{L_c(z)}\right)^{\beta} \right\}^{-1}\,.
\end{equation}
\begin{table}[h]
\centering{
\begin{tabular}{|c|c|c|c|c|c|}
\hline
$\rho_0$&$\alpha$&$\beta$ & $a_0$   &  $a_1$	&  $a_2$	 \\ \hline \hline
 $10^{0.85}$ Gpc$^{-3}$ &$0.83$ &$1.96$  & $24.89$ &$1.18$  &  $-0.28$  \\ \hline
\end{tabular}
} 
\caption{Parameters for the flat spectrum RLF [\cite{peacock}].}
\label{peacock_params}
\end{table}
\section{The single source spectrum \label{single_source}}
The generic AGN neutrino flux is adopted to follow the proton spectrum
from the $\Delta$-decay.
This can be described as a power law with an index $p$ having an exponential
cutoff. This cutoff depends on the particle's maximum energy. Also, the power of the spectrum changes by $-2$ when synchrotron
losses of pions start to dominate at energies of $E_{\nu}^{break}\approx
E_{\nu}^{\max}/6.7$. The break energy is determined by the ratio of the pion's
to the protons mass, $m_{\pi}/m_{p}= 1/6.7$, since the maximum energy due to
synchrotron losses of a charged particle is
proportional to the particles mass~[\cite{biermann_strittmatter}]. The spectrum is then given as
\begin{equation}
\frac{dN}{dE_{\nu}} = \left\{
\begin{array}{lll}
\phi _{\nu }\left(L_{disk}\right) \cdot E_{\nu }^{-p}
  \exp\left(-\frac{E_{\nu}}{E_{\nu}^{\max}}\right) &&\mbox{for }
  E_{\nu}<E_{\nu}^{break}\\
\phi _{\nu }\left(L_{disk}\right) \cdot E_{\nu }^{-p}
  \left(E_{\nu}/E_{\nu}^{break}\right)^{-2} \exp\left(-\frac{E_{\nu}}{E_{\nu}^{\max}}\right) &&\mbox{for } E_{\nu}\geq E_{\nu}^{break}
\end{array}
\right.\,.
\end{equation}
The power law
behavior is due to diffuse shock acceleration of the protons in the jet's shock fronts. The spectrum is limited by the strength of the
magnetic field of the source due to the spatial limit of the Larmor motion~[\cite{hillas}]. This determines the maximum energy of
the accelerated protons, there is an exponential cutoff in the spectrum. The
magnetic field is connected to the disk luminosity of the AGN and subsequently
the maximum proton energy can be expressed in terms of the disk luminosity, see~[\cite{lovelace}]:
\[
E_{p}^{\max}\propto\sqrt{L_{disk}}
\] 
as it has been confirmed by the jet-disk symbiosis model,
see~[\cite{falcke1}]. At high luminosities of $L_{disk}>10^{46}$~erg/s \cite{biermann_erice}, a
saturation of the proton maximum energy is expected due to photo-pion
production in the disk radiation field. This effect will be discussed
in section~\ref{limits}.

The normalization of the generic neutrino spectrum, $\phi _{0 }$,
can be found by assuming that particles from an AGN produce the same 
order of magnitude of luminosity as the jet power,
$Q_{jet}$, and neutrinos contribute with a fraction $q_{\nu}< 1$~[\cite{falcke}]. 
The observations are best fitted by this concept and this number constraint.
It is assumed that the power of the
neutrinos is directly connected to the power of the jet. This in turn is
a fraction $q\geq 1/3$ of the disk luminosity. Here we use the lower limit,
$q=1/3$, as an estimate. 
The factor $q_{\nu}$ results from assuming a kinematic factor of $\xi_{kin}\approx 1/2$ which  accounts for the fact that only half of the protons are converted into pions - about 50\% of the protons convert into neutrons~[\cite{muecke2000}]. Additionally, a (non
thermal) efficiency of $\epsilon_{nth}\approx 10\%$ is applied - 20\% of the proton energy goes into secondary mesons and 
50\% of these 20\% go into neutrino production. The optical depth of AGN
sources $\tau_{p\gamma}$ can be described as the product of the photohadronic
cross section $\sigma_{p\gamma}$ and the photon density $n_{\gamma}$, and thus
\[
\tau_{p\gamma}\approx 800\,\frac{L_{46}}{r_{17}}\,.
\]
Here, $r_{17}:=r/10^{17}$~cm is the extension of the source in units of
$10^{17}$~cm. Scaling this optical depth to the Eddington luminosity and to the
Schwarzschild radius demonstrates that the optical depth is independent of the 
power of the source at equivalent distances from the central engine.  It
also shows that large optical depths are easily achieved. 
The neutrino measurements can then be used to set limits on the optical depths, and so
on the structure of the emitting region, as we show below.
Taking into account interactions of protons in Bethe-Heitler pair production, an 
effective opacity for photopion production is reduced to 
\[
\tau_{eff}=\tau_{p\,\gamma}\cdot\eta\,.
\]
$\eta$ gives the
reduction of the optical depth due to the optical depth of the source taking
into account Bethe Heitler interactions, $\tau_{BH}$. In all following
calculations, $\tau_{eff}=1$ will be assumed. Detailed geometric models for emission from zones where the
effective optical depth is large, where it is close to unity, and where it
is much less than unity, need to be considered in a later stage of such
calculations. This might be easier in the future once there are good constraints, or even
 exact spectra from multi-GeV to multi-TeV observations of AGN.
Since the neutrino flux
is proportional to $\tau_{eff}$, a lower limit as
an estimate of a maximal optical depth is possible when comparing current
neutrino flux limits to the calculated flux as it is described in section \ref{final_spectra}.

The fraction $q_{\nu}$ is then
\[
q_{\nu}=q\cdot\tau_{eff}\cdot \xi_{kin}\cdot\epsilon_{nth}\approx\frac{1}{60}\,.
\]
Thus the neutrino luminosity
from the jets is given as a fraction of the disk luminosity:
\begin{eqnarray*}
\int_{E_{\nu}^{\min}}^{E_{\nu}^{\max}}E_{\nu } \cdot \frac{dN}{dE_{\nu}} \cdot
dE_{\nu} &=& \int_{E_{\nu}^{\min}}^{E_{\nu}^{break}}\phi _{\nu } \cdot
E_{\nu}^{-p+1}\exp\left(-\frac{E_{\nu}}{E_{\nu}^{\max}} \right) dE_{\nu }\\
&+&\int_{E_{\nu}^{break}}^{E_{\nu}^{\max}}\phi _{\nu }\cdot E_{\nu
}^{-p-1}\cdot {E_{\nu}^{break}}^{2}\,\exp\left(-\frac{E_{\nu}}{E_{\nu}^{\max}} \right) dE_{\nu }
= q_{\nu} \cdot L_{disk} \,.
\end{eqnarray*}
The integral can be evaluated by developing the exponential function in a Taylor series and interchanging
the integral with the sum. 
The lower integration limit is a fraction of $1/4$ of the rest mass of the
pion, $E_{\nu}^{\min}=E_{\pi^{\pm}}^0/4 = (139.57018\pm0.00035)/4 \mbox{ MeV}$, [\cite{ppb}]. The upper integration limit is
assumed to be varying with the disk luminosity, see
Equ.~(\ref{love_emax}). 

The normalization is thus given as
\begin{eqnarray*}
x\,L_{disk}&=&\ln\left(\frac{E_{\nu}^{\max}}{E_{\nu}^{\min}}
\right)+\left\{f(p-2,1/6.7,0)\right.\\[0.1cm]
&+&\left.\left[f(p,1,0)-f(p,6.7,0)\right]/6.7^2\right\}\cdot{E_{\nu}^{\max}}^{-p+2}\\[0.1cm]
&-&f(p-2,E_{\nu}^{\max}/E_{\nu}^{\min},-p+2)\cdot {E_{\nu}^{\min}}^{-p+2}
\end{eqnarray*}
with
\[
f(a,b,c)=
\sum_{n=0,n\neq a}^{\infty} \frac{(-1)^n}{n!\cdot
  (n-a)}b^{-n+a+c}\,.
\]
The series' are evaluated numerically for each case.
\subsection{Jet-disk symbiosis}
To convert the disk luminosity into the radio luminosity of the jets, the jet-disk symbiosis
model by \cite{falcke} is used. 
\subsubsection{Flat spectrum sources}
Assuming a power law slope of $\alpha=0$ the luminosity of the
jet in units of erg/s, $L^{f}$ is given as [\cite{falcke1}]
\begin{equation}
\begin{array}{c}
L^{f}= 6.7\cdot10^{42}\frac{erg}{s} \cdot 
D^{2.17}\cdot \sin^{0.17}(i_{obs})\cdot \left(x_e' \right)^{0.83}\\[0.3cm]
\cdot
\left(\frac{6}{\gamma_j} \right)^{1.8} \cdot q_{j/1}^{1.42}\cdot L_{46}^{1.42-\xi}\,.
\end{array}
\label{falckecomp}
\end{equation}
Parameters appearing in Equ.~\ref{falckecomp} are given in the parameter list below.
In using here $\alpha = 0$ instead of the sum of components with $\alpha =
1/2$ we follow the earlier work. It is not known what the energetic particle spectrum 
does at low energies in the jet. For the final conclusion below we do use $p = 2$, which would correspond to $\alpha = 1/2$ if pure synchrotron radiation was observed.

The parameters from above are defined in the parameter list below. 
The correlation between disk luminosity and radio luminosity in units of $10^{42}$~erg/s, $L_{42}^{f}$, is 
\begin{equation}
L_{disk}=(2.1\pm 1.9) \cdot 10^{45} (L_{42}^{f})^{0.79} \left[ \frac{erg}{s}\right]\,.
\end{equation}
The error has been estimated by taking into account a scattering of the data
by about a factor three around $L_{42}^{f}(L_{disk})$.
\subsubsection{Steep spectrum sources}
The model for extended, optically thin sources will additionally be
used for the calculation of the steep spectrum source flux. The relation between the differential radio luminosity 
at $5$~GHz of the lobes and the disk luminosity is
\begin{equation}
P_{lobe}=1.8\cdot 10^{34} \frac{erg}{s\cdot Hz}\left(\frac{GHz}{\nu} \right)^{0.5} \cdot 
\frac{\beta_j^{1/4}\cdot P_{-12}^{1/4}\cdot x'_{e,100}\cdot \left(q_{j/1}L_{46} \right)^{3/2}}{\gamma_{j,5}^{7/4}\cdot u_3^{3/4}}\,.
\label{falckeext}
\end{equation}
This result is based on the investigation of a quasar sample,
described in detail in~[\cite{falcke1}]. The disk 
luminosity is connected to the integral radio luminosity, $L^{s}$, as
\begin{equation}
L_{disk}=(21.6\pm 16.6)\cdot 10^{45} \cdot (1+z)^{-1/2}(L_{42}^{s})^{2/3}\,,
\end{equation}
assuming a spectral index of $\alpha=0.8$. Again, the error has been estimated
considering the scattering of the data by a factor three around the predicted result.
The parameters which occur in Equ.~(\ref{falckecomp}) and~(\ref{falckeext}) are given below.

\begin{small}
PARAMETER LIST
\begin{itemize}
\item $\beta_j$: Jet velocity in units of $c$ which can be assumed to have a value of $\beta_j\approx 0.986$.
\item $\gamma_j$: jet's Lorentz Boost factor,  $\gamma_j=6$.
\item $P_{-12}$: Pressure parameter. The external radiation pressure at the AGN is given by $P_{ext}=P_{-12}\cdot 10^{-12}\mbox{ erg/cm}^3$. The pressure evolves with the time due to the expansion of the Universe. The pressure parameter depends on the redshift as $P_{-12}=(1+z)^3\cdot P_{-12}^0$, where  $P_{-12}^0$ is the pressure of the Universe at $z=0$. $P_{-12}^0$ is set to $P_{-12}^0=1$.
\item $x_{e,100}'$: Modified electron density. Let the ratio between the relativistic electron
density and the total number density of protons be $x_e$ and the
minimum Lorentz factor of the relativistic electron population divided
by 100, $\gamma_{e,100}$. The modified electron density is then $x_{e,100}'=\gamma_{e,100}^{p-1}x_e$. $p=2$ is a parameter. $x_{e,100}'\approx 1$ has been used in all following calculations~[\cite{falcke1}].
\item $q_{j/1}$: Total jet power $Q_{jet}$ in units of the disk luminosity, with a given value of $q_{j/1}=Q_{jet}/L_{disk}=0.15^{+0.2}_{-0.1}$.
\item $L_{46}$: Disk luminosity per $10^{46}\mbox{ erg/s}$: $L_{46}=L_{disk}/(10^{46}\mbox{ erg/s})$. 
\item $\gamma_{j,5}$: Jet's Lorentz boost factor, divided by 5. It is assumed to be $\gamma_{j,5}=(6\pm2)/5$.
\item $u_3$: Ratio between the total energy density in the jet and the magnetic energy density, divided by a factor three. It is set to $u_3=1$ [\cite{falcke}].
Note, again, that this fits the observations best.
\item $\xi=0.15$: Fit parameter resulting from the ansatz that the bulk proper
  velocity depends on the accretion rate and thus on the disk luminosity, see [\cite{falcke1}].
\item $i_{obs}=1/\gamma_j=1/6$: angle between observer and jet axis. The angle is taken to be $1/\gamma_j$, because this is the 
maximal boosting range and has a solid angle in which optimal boosting
occurs. We use for simplicity the same angle for all
sources: In a more refined model obviously isotropy should be used, but
then mediated by the selection effects, which strongly favor small
angles, as are used here.
\item $D=\frac{1}{\gamma_j}\,(1-\beta_j\,\cos i_{obs})^{-1}\approx 6$: Doppler factor. In this calculation, the Doppler
factor is assumed to be the same for all sources. A variation of the Doppler factor with the source luminosity has been discussed - 
for an example see e.g.~[\cite{falcke}]. However, since there is
no well-founded model to apply individual Doppler factors so that a constant factor has to be used as an approximation.

\end{itemize}
\end{small}

\begin{figure}
\setlength{\unitlength}{1cm}
\begin{minipage}[t][10cm][b]{6.7cm}
\begin{picture}(6.7,3.5)
\includegraphics[width=7.5cm]{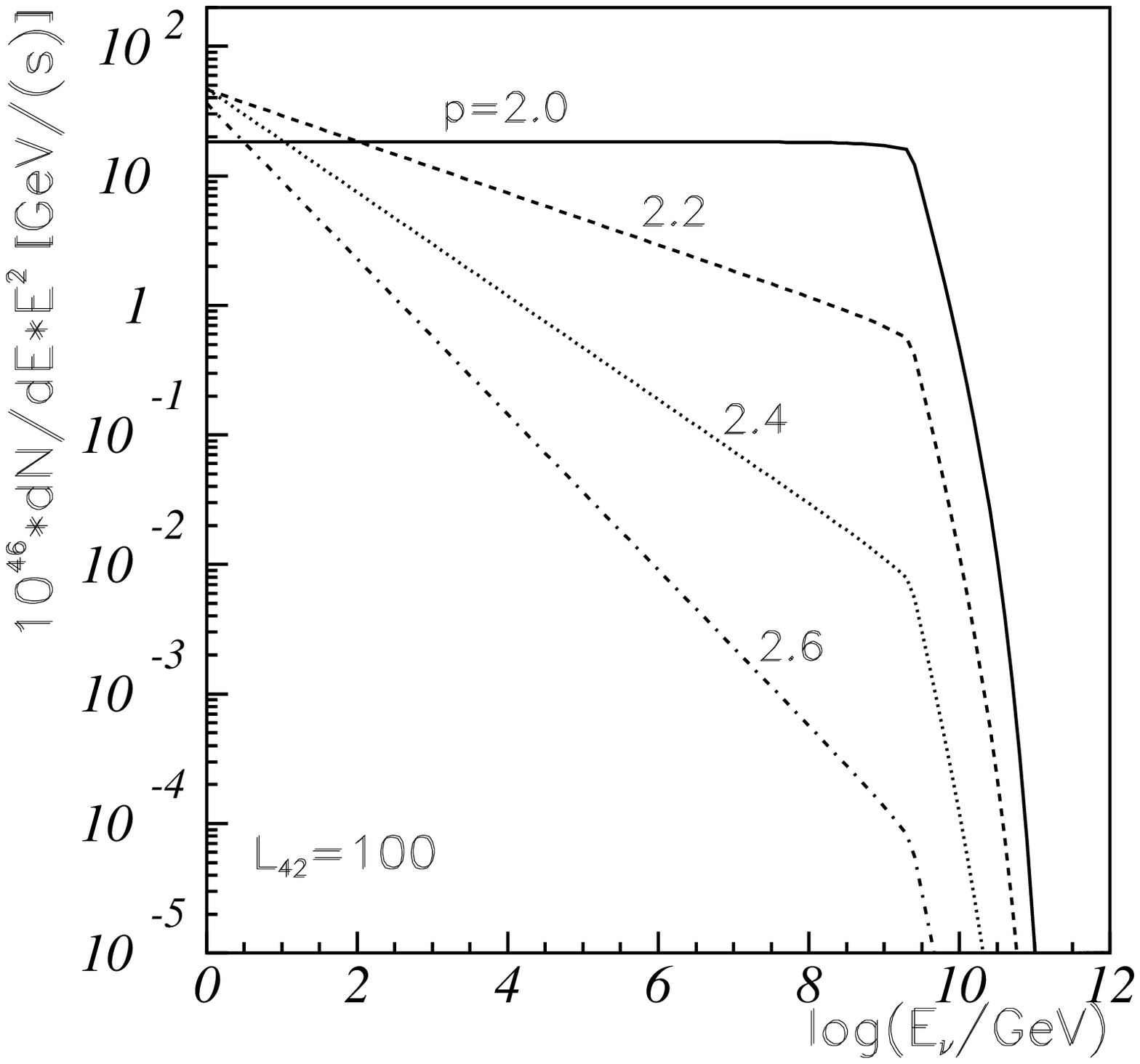}
\end{picture}\par
\caption[Generic AGN neutrino
  spectrum of steep spectrum sources.]{Energy dependence of the generic AGN neutrino
  spectrum of steep spectrum sources. The normalization is dependent on the particle index $p$, shown for $p=2.0,\,2.2,\,2.4,\,2.6$.
The radio luminosity is assumed to be $L_{42}=100$.}
\label{endep_s_paper}
\end{minipage}
\parbox{0.5cm}{\quad}
\begin{minipage}[t][10cm][b]{6.7cm}
\begin{picture}(6.7,3.5)
\includegraphics[width=7.5cm]{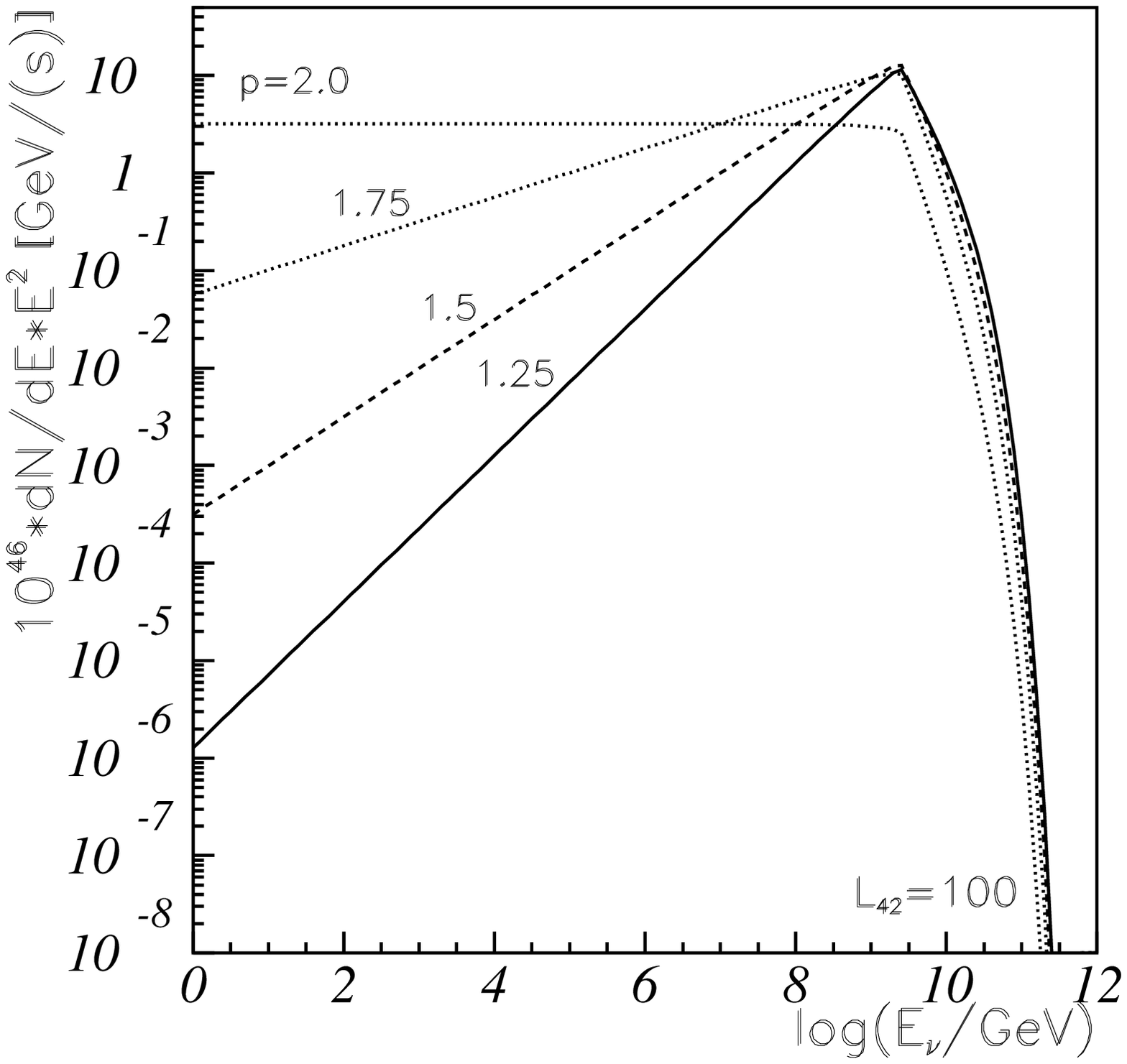}
\end{picture}\par
\caption[Generic AGN neutrino spectrum of flat spectrum sources.]{Energy dependence of the generic AGN neutrino
  spectrum flat spectrum sources. The spectrum is shown at $L_{42}=100$ and for three different particle indices, 
$p=1.25,\,1.50,\,,\,1.75,\,2.0$.\\[0.cm]}
\label{endep_f_paper}
\end{minipage}\hfill
\end{figure}

Finally, the jet-disk symbiosis can be used to express the generic neutrino
energy spectrum in terms of the radio luminosity in units of
$10^{42}$~erg/s.
The result at a luminosity of $L_{42}=100$ with varying spectral index is shown for steep spectrum sources
in Fig.~\ref{endep_s_paper} and in Fig.~\ref{endep_f_paper} for flat spectrum sources. 
The power law decrease can be observed up to a cut energy at $E_{\nu}\sim 10^{9}$~GeV, where synchrotron losses of the
neutrino-producing pion start to show as well as the exponential cutoff of the spectrum.

~\vspace{0cm}
\section{Integration limits \label{limits}}
\begin{figure}[ht]
\setlength{\unitlength}{1cm}
\begin{minipage}[t][9.5cm][b]{6.7cm}
\begin{picture}(6.7,3.5)
\includegraphics[width=7.5cm]{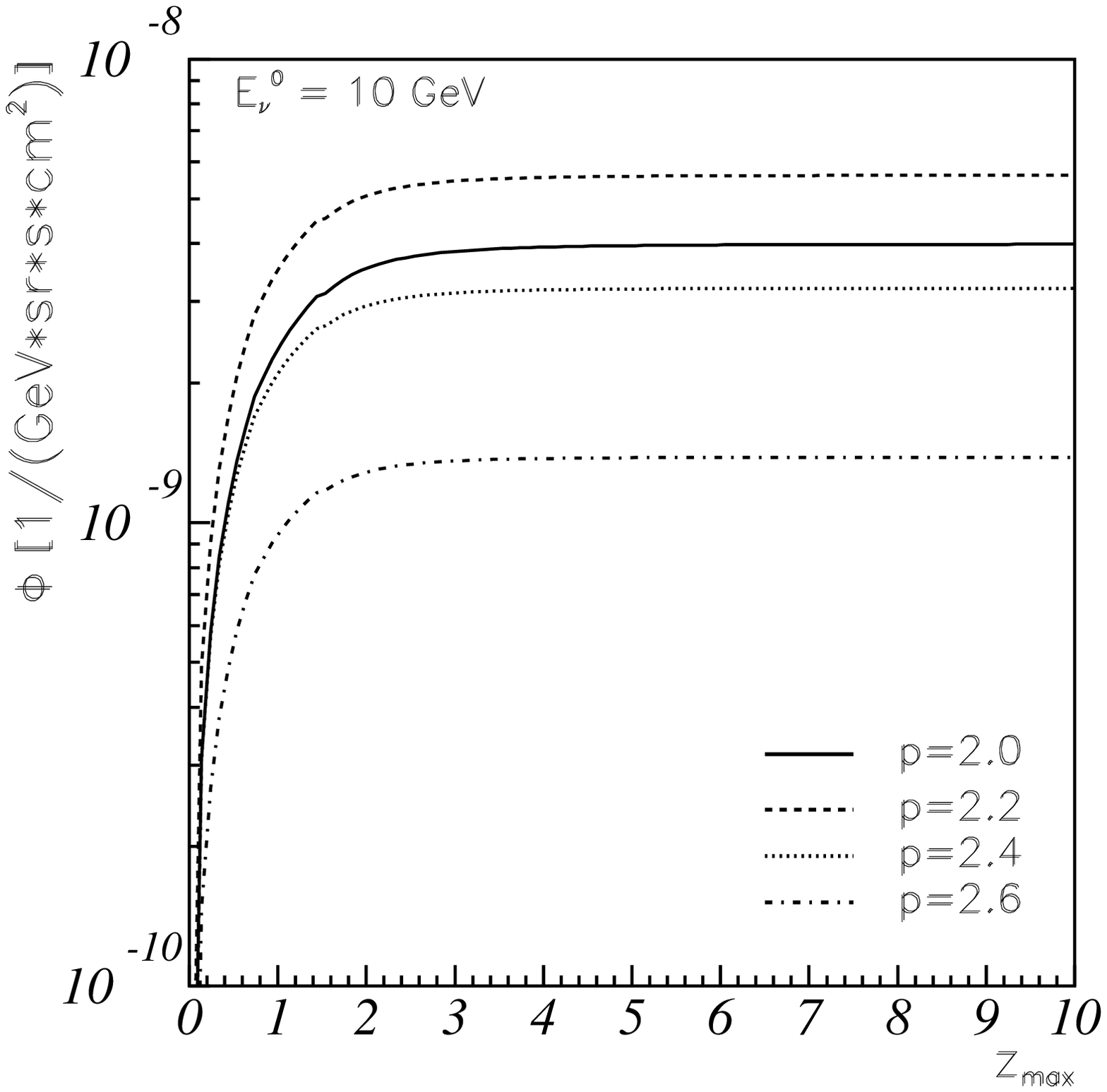}
\end{picture}\par
\caption{Neutrino flux with variation of
the upper redshift integration limit $z_{\max}$ for steep spectrum sources. The neutrino energy is set to $E_{\nu}^{0}=10$~GeV while the particle index is varied as $p=2.0,2.2,2.4,2.6$.}
\label{willott_paper_zmax}
\end{minipage}
\parbox{0.5cm}{\quad}
\begin{minipage}[t][9.5cm][b]{6.7cm}
\begin{picture}(6.7,3.5)
\includegraphics[width=7.5cm]{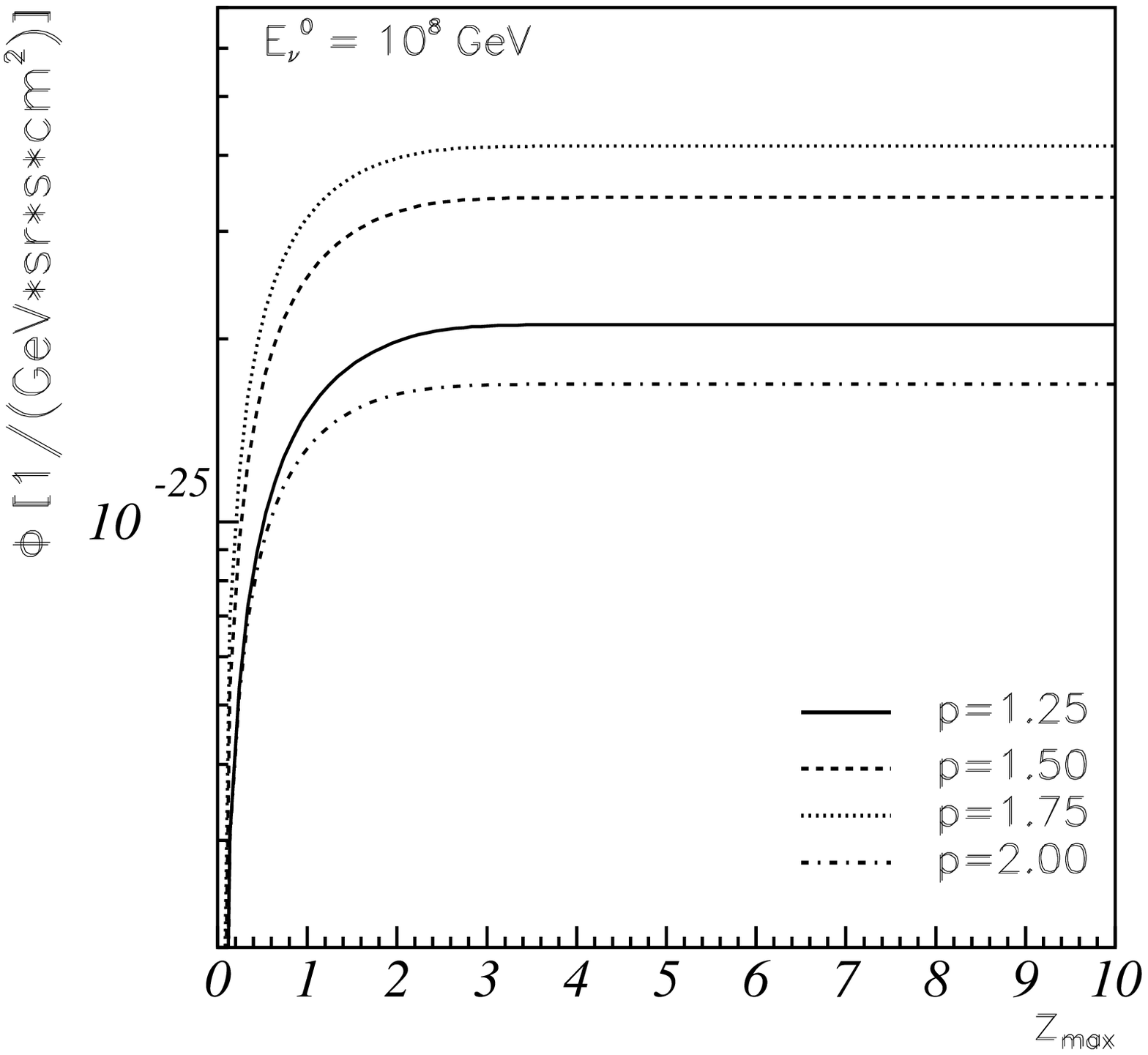} 
\end{picture}\par
\caption{Neutrino flux with variation of
the upper redshift integration limit $z_{\max}$ for flat spectrum sources. The neutrino energy is set to $E_{\nu}^{0}=10^8$~GeV while the particle index is varied as $p=1.25,1.50,1.75,2.00$.}
\label{peacock_paper_zmax}
\end{minipage}\hfill
\end{figure}

The integration over the luminosity has to be performed before the redshift integration, since the lower luminosity limit implicitly depends on the redshift. The
limits for the z-integration are taken to be 
\begin{eqnarray*}
z_{\min}&=&0.03\\
z_{\max}&=&6 \,.
\end{eqnarray*}
The minimum of the $z$-integration is given by the fact that the flux is assumed to be
diffuse. For $z<0.03$, a contribution from the
supergalactic plane is expected as will be discussed in section~\ref{results}.
The maximum redshift is
taken to be $z=6$, since the total z dependence decreases rapidly with
redshift and any contribution above $z_{\max}$ can be neglected as 
can be seen in Fig.~\ref{willott_paper_zmax} (steep spectrum
sources) and Fig.~\ref{peacock_paper_zmax} (flat spectrum sources).
Up to $z=6$, the consistency of the used models is
ensured, since AGN have been detected up to redshifts of $z\approx 6.4$ [\cite{willott_z}].

\begin{figure}[ht]
\setlength{\unitlength}{1cm}
\begin{minipage}[t][11.5cm][b]{6.7cm}
\begin{picture}(6.7,3.5)
\includegraphics[width=7.5cm]{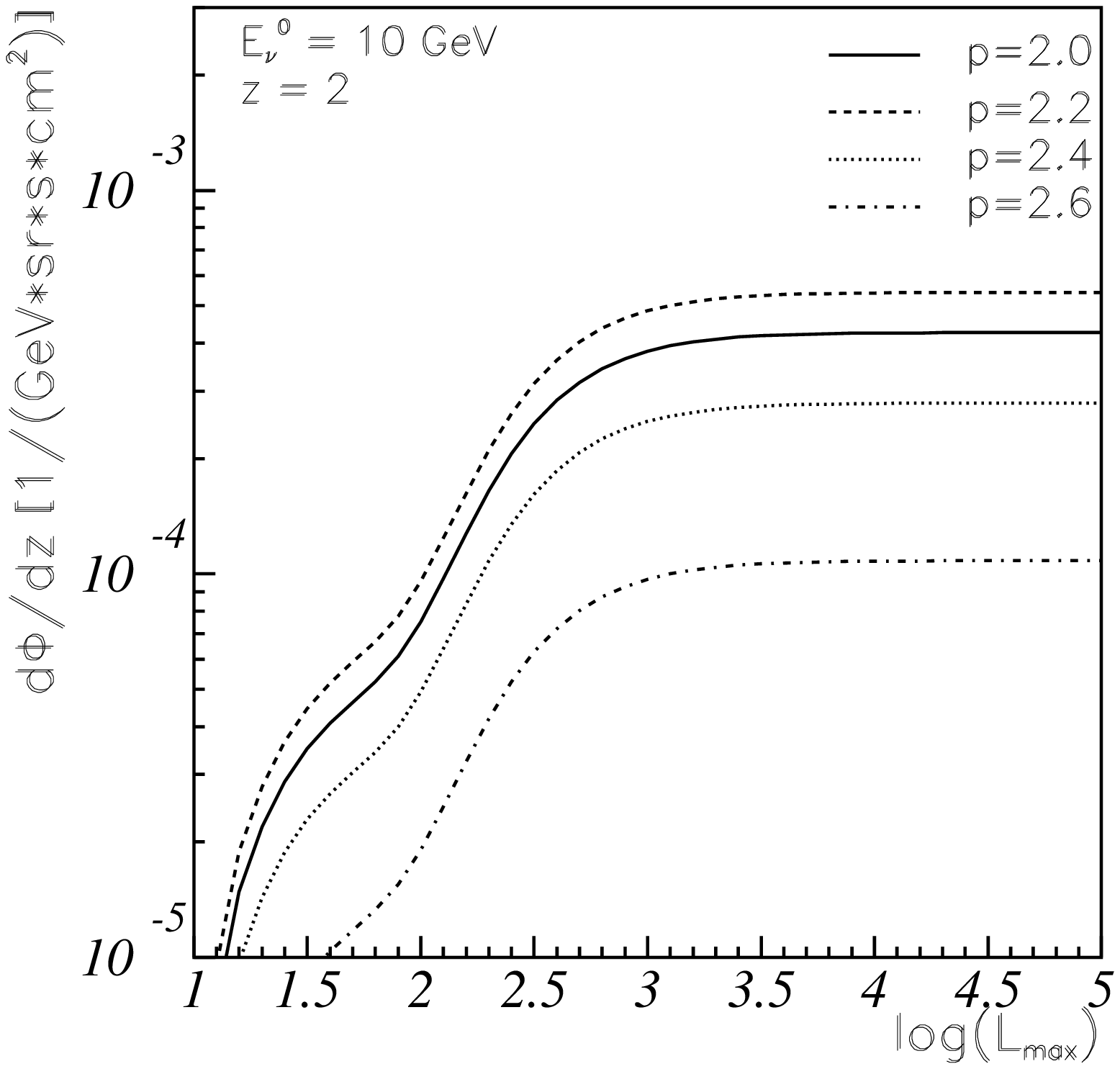}
\end{picture}\par
\caption{Variation of the steep spectrum source neutrino flux with the upper integration limit, 
using different spectral indices, $p=2.0,\,2.2,\,2.4$ and $2.6$. The redshift is arbitrarily taken to be $z=2$, 
the energy is set to $E_{\nu}^{0}=10$~GeV. The flux normalization is saturated at approximately $L_{\max}\approx 10^{3}$.}
\label{willott_paper_lmax}
\end{minipage}
\parbox{0.5cm}{\quad}
\begin{minipage}[t][11.5cm][b]{6.7cm}
\begin{picture}(6.7,3.5)
\includegraphics[width=7.5cm]{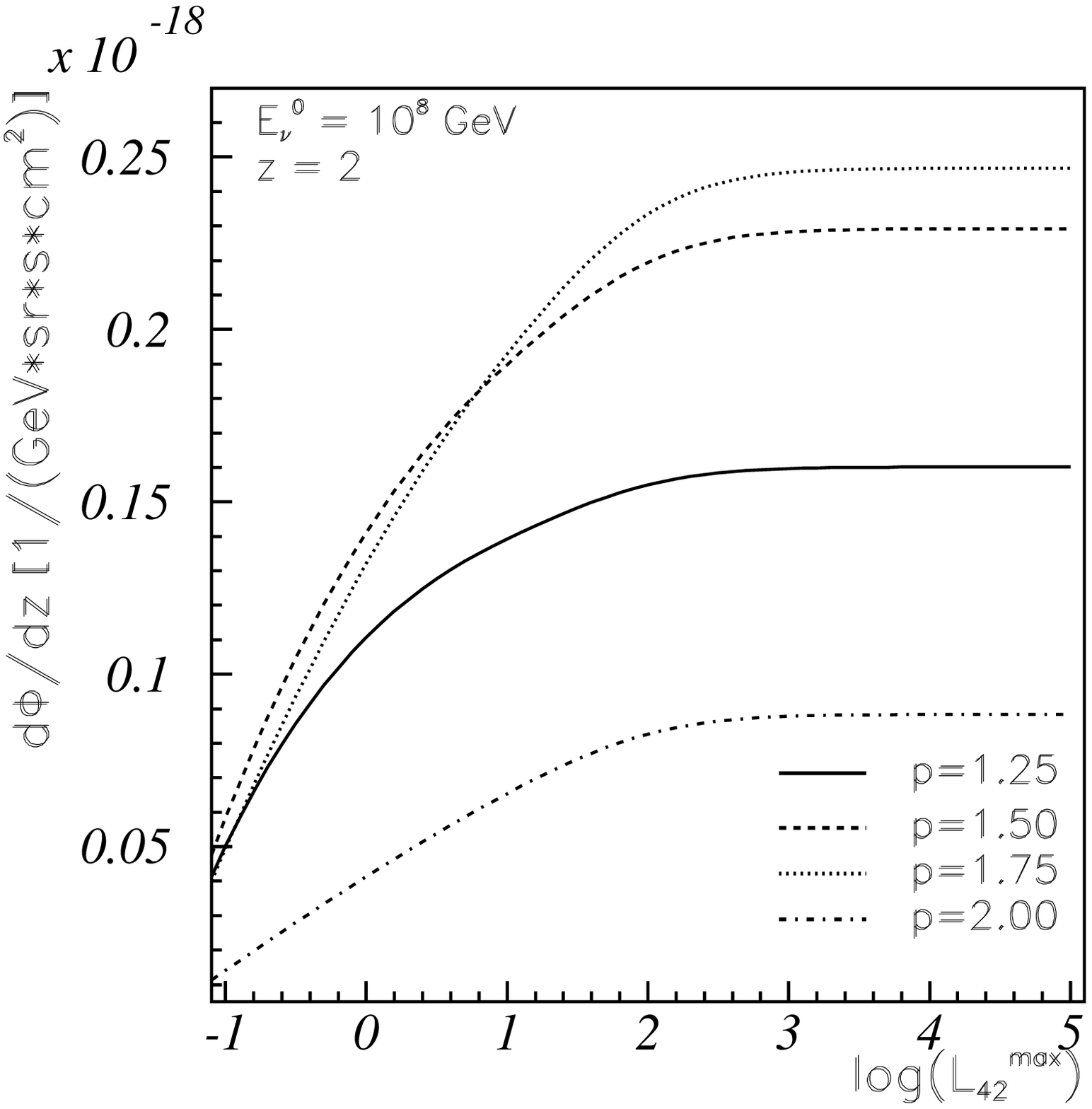}
\end{picture}\par
\caption[]{Variation of the neutrino flux (flat spectrum sources), using $p=1.25,\,1.5,\,1.75$ and $2.00$. The redshift is chosen to be $z=2$ and the energy is set to $E_{\nu}^{0}=10$~GeV. Sources at $L>10$ do not contribute significantly to the flux.\\[0.0cm]}
\label{peacock_paper_lmax}
\end{minipage}\hfill
\end{figure}

The maximum radio luminosity limit is 
taken to be $L_{42}^{\max}=10^{4}$.  
Fig.~\ref{willott_paper_lmax} (steep spectrum sources) and
\ref{peacock_paper_lmax} (flat spectrum sources) show the dependence of the flux at $z=2$ and $E_{\nu}^{0}=10^8$~GeV (flat) 
respectively $E_{\nu}^{0}=10$~GeV. The contribution is negligible for $L_{42}>10^{3}$ 
in the case of steep spectrum sources and for $L_{42}>10$ in the case of flat spectrum sources.
This shows that most of the contribution to the flat spectrum neutrino flux
comes from sources with moderate luminosities, $L_{42}<10$. 
The highest contribution to the flat spectrum neutrino flux comes from sources with luminosities around $L_{42}\approx 300-1000$.

The lower radio luminosity limit depends on the following
facts:
\begin{enumerate}
\item The absolute lower limit: For the model of \emph{extended sources}, the
jet-disk symbiosis is developed for FR-II galaxies and the
absolute lower limit, $L_{42,\min}^{s}$, is given by the lower luminosity limit for FR-II
galaxies which is given as $L_{0.178}\approx 2.5\cdot 10^{26}$~W/Hz
with an integral radio luminosity of approximately $L_{42,\min}^{s}=10$.
The sample of \emph{flat spectrum sources}, on the other hand, ranges from
$10^{-1.5}<L_{42}^{f}<10^{4}$ and the absolute lower limit, $L_{42,\min}^{f}$ is thus
$L_{42,\min}^{f}=10^{-1.5}$. 
\item The maximum proton energy which can be produced by an AGN is
  connected with the disk luminosity according to~\cite{lovelace}:
\begin{equation}
E_{p}^{\max}=C'\cdot \sqrt{L_{46}}
\label{love_emax}
\end{equation}
up to a luminosity of $L_{46}\approx 1$, where the relation saturates,
\begin{equation}
E_{p}^{\max}=C'\,.
\end{equation}
The result is supported by the jet-disk symbiosis model, see~[\cite{falcke1}]. 
Taking the highest observed energy so far, $E_p\approx 10^{21}\mbox{eV}$ and the maximum luminosity, $L_{disk}^{\max}=10^{47}$ erg/s,
  the constant $C'$ is determined to be $C'=10^{11.5}\mbox{GeV}$. 
When  the jet-disk symbiosis model by Falcke et al. is used and the
  ratio between $E_p$ and $E_\nu$ taken to be $20/1$, the maximum
  neutrino energy is correlated to the radio luminosity due to the non linear
  relation between radio and disk luminosity,
\[
E_\nu^{\max}=C\cdot L_{42}^{\beta}\,.
\]
This numerical relation is consistent with inserting all known parameters for the 
radio galaxy M87. $C$ and $\beta$ are given by the corresponding relations between disk
and radio luminosity for flat or steep spectrum sources
and are listed in table~\ref{emax_params}.

\noindent Therefore, the lower luminosity limit of an AGN contributing to a
certain flux at a fixed neutrino energy is
\begin{equation}
L_{42}^{\min}=\max\{L_{42,\min}^{s/f},\left(\frac{E_\nu}{C}\right)^{1/\beta}\}\,.
\end{equation}
The contribution to the flux for an energy corresponding to a luminosity
exceeding $L_{disk}=10^{46}$~erg/s, that is energies of
$E_{\nu}=C'/20=10^{11.5}$~GeV$\approx 2\cdot 10^{10}$~GeV is insignificantly small. 
\end{enumerate}
\begin{table}
\centering{
\begin{tabular}{|c|c|c|c|}
\hline
&$C$ [GeV]&$\beta$&$E_{\nu}^{cut}$ [GeV]\\\hline \hline
steep&$2.2\cdot 10^{10}\cdot (1+z)^{-1/4}$&$1/3$& $\sim 5\cdot 10^{9} (z=2)$\\ \hline
flat&$7.5\cdot 10^{9}$&$0.3935$&$\sim 2\cdot 10^{8}$\\ \hline
\end{tabular}
}
\caption{Parameters for the determination of the lower luminosity
integration limit.}
\label{emax_params}
\end{table}
Since the RLF is decreasing with luminosity, the energy spectrum will
steepen from that energy on, when the luminosity-energy relation
exceeds the value of the absolute luminosity limit,
$\left(\frac{E_\nu}{C}\right)^{1/\beta}\geq L_{42,\min}^{f/s}$. This is
the case for an energy 
\begin{equation}
E_\nu^{cut}/\mbox{GeV} \approx (L_{42,\min}^{s/f})^{\beta}\cdot C\mbox{/GeV}
\end{equation}
in the comoving frame. The cut energies for the two models are given
in table~\ref{emax_params}.
\section{Results\label{results}}
\subsection{Discussion of the spectra varying the particle index}
\begin{figure}[ht]
\setlength{\unitlength}{1cm}
\begin{minipage}[t][10cm][b]{6.7cm}
\begin{picture}(6.7,3.5)
\includegraphics[width=7.5cm]{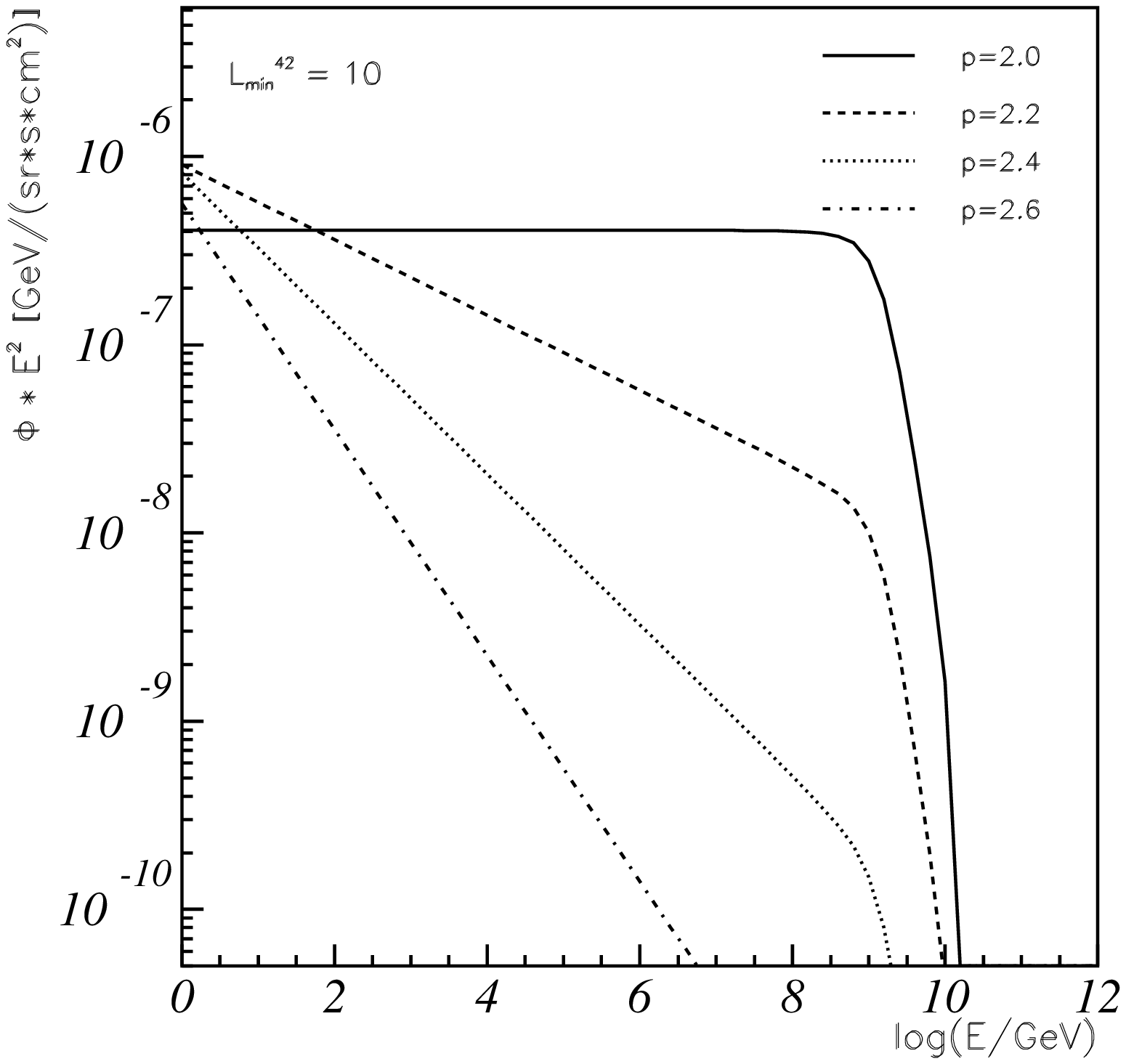}
\end{picture}\par
\caption[Neutrino spectrum of steep spectrum AGN.]{Neutrino spectrum of steep spectrum AGN. The particle index is varied ($p=2.0,2.2,2.4,2.6$).}
\label{flux_final_paper}
\end{minipage}
\parbox{0.5cm}{\quad}
\begin{minipage}[t][10cm][b]{6.7cm}
\begin{picture}(6.7,3.5)
  \includegraphics[width=7.5cm]{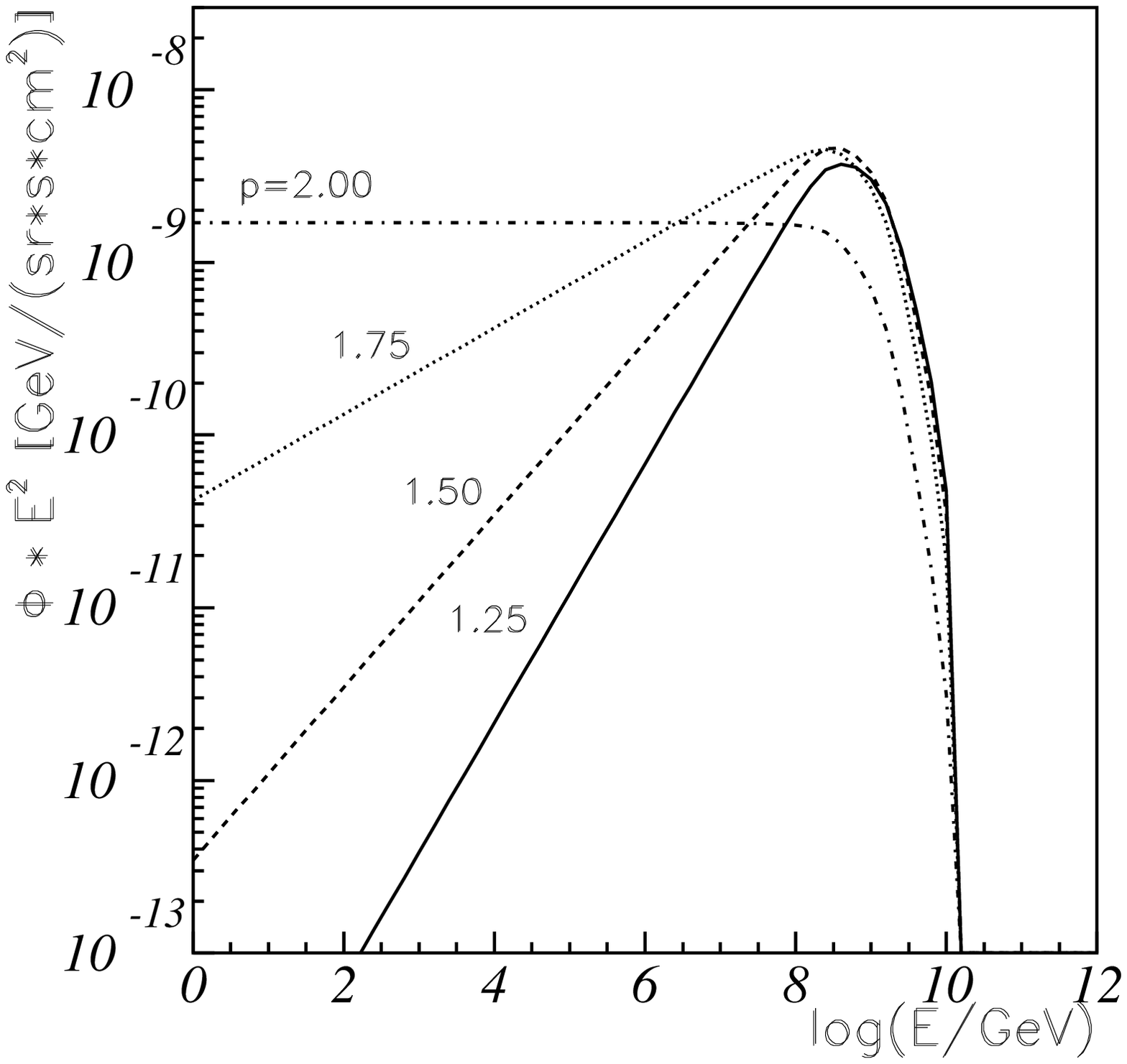}
\end{picture}\par
\caption[Neutrino spectrum of flat spectrum AGN.]{Neutrino spectrum
  of flat spectrum AGN for different particle indices $p=1.25,1.50,1.75,2.00$.}
\label{peacock_paper}
\end{minipage}\hfill
\end{figure}
The final steep spectrum source flux is shown in Fig.~\ref{flux_final_paper}. 
The spectrum is presented for four different particle indices, $p=2.0,2.2,2.4,2.6$. 
Here, $p=2$ is a commonly used value in literature and $p=2.6$ has been computed in section~\ref{samples} from the 
spectral index at $2.7-5$~GHz for the steep spectrum population. The spectrum decreases as a power law up to a cut energy at
approximately $10^{9}$~GeV. At higher energies, the spectrum is falling more
rapidly due to pion synchrotron loss effects and the exponential cutoff of the
spectrum. For a particle index of $p=2$, 
the flux is approximately 
\[
E^2\,\Phi^{s}\approx 10^{-7}\mbox{GeV/}(\mbox{s sr cm}^2)
\]
for one neutrino flavor.
This is very close to the current neutrino flux
limit derived from the observed atmospheric muon and anti muon neutrino spectrum 
\[
E^2\,\Phi < 2.6\cdot 10^{-7} \mbox{GeV/}(\mbox{s sr cm}^2)
\]
at energies around 300~TeV [\cite{AMANDA}]. 
The flat source neutrino spectrum is shown in Fig.~\ref{peacock_paper} with a spectral 
index of $p=1.25,1.50,1.75$ and $2.00$. The cut energy lies at $E\sim
10^{9}$~GeV as discussed before. Values $p\neq2$ are shown as a comparison,
the final result will be given for $p=2$ as indicated by diffuse shock
acceleration. For $p=2$, the flux is
\[
E^2\,\Phi^{f}\approx 10^{-9.2}\mbox{GeV/}(\mbox{s sr cm}^2)
\]
for one neutrino flavor. This is about two orders of magnitude lower than the
prediction for steep spectrum sources.
\begin{figure}
\setlength{\unitlength}{1cm}
\begin{minipage}[t][13.5cm][b]{6.7cm}
\begin{picture}(6.7,3.5)
\includegraphics[width=7.5cm]{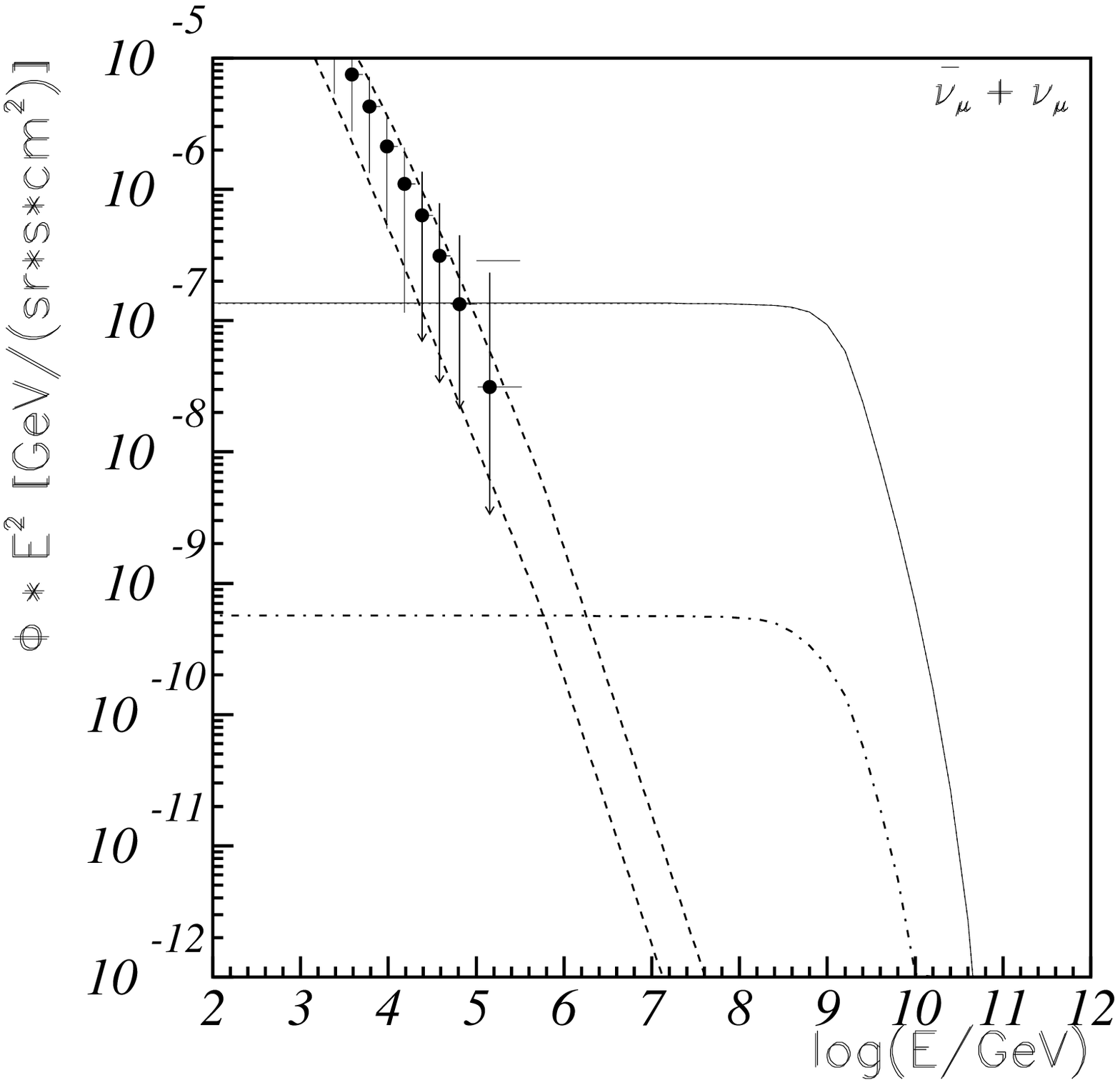}
\end{picture}\par
\caption[AGN neutrino spectrum with an index of $2$.]{AGN
  $\nu_{\mu}+\overline{\nu}_{\mu}$ neutrino spectrum with an index of $p=2$. 
The flat spectrum sources (dotted line) do not contribute 
significantly compared to the steep spectrum population - 
the solid line represents the sum of flat and steep spectrum sources which
does not differ from the result for steep spectrum sources. 
The data points result from unfolding the AMANDA neutrino
spectrum. They follow the conventional atmospheric neutrino spectrum (dashed lines) and an upper limit is derived 
as is indicated
in the figure~[\cite{AMANDA}].}
\label{flux_2_paper}
\end{minipage}
\parbox{0.5cm}{\quad}
\begin{minipage}[t][13.5cm][b]{6.7cm}
\begin{picture}(6.7,3.5)
\includegraphics[width=7.5cm]{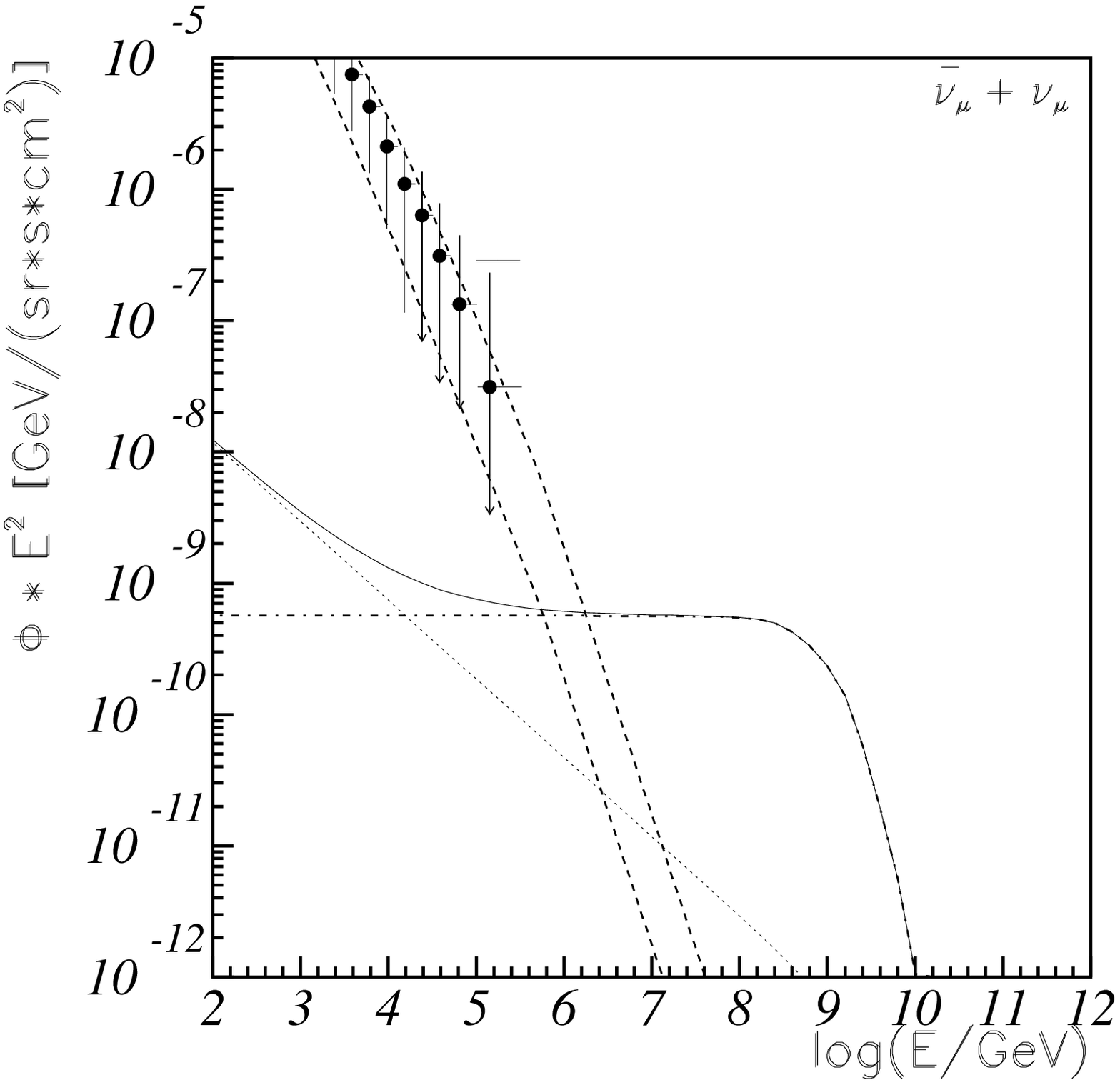}
\end{picture}\par
\caption[]{AGN $\nu_{\mu}+\overline{\nu}_{\mu}$ neutrino spectrum with an index
  of $p=2.6$ for steep spectrum sources and $p=2.0$ for flat spectrum sources. 
The steep spectrum sources (dot-dashed line) make up most of the spectrum at lower energies ($E_{\nu}>10^{6}$~GeV). 
The flat spectrum sources (dotted line) dominate the total spectrum (solid line)
at energies of $E_{\nu}>10^{6}$~GeV. A signal that exceeds the atmospheric 
contribution is expected starting at $E_{\nu}\approx 10^{8}$~GeV.\\[0.95cm]}
\label{flux_26}
\end{minipage}\hfill
\end{figure}
\subsection{Discussion of Uncertainties \label{uncertainties}}
Apart from the lack of knowledge of the exact optical depth, the calculation of the neutrino flux $\Phi$ bears uncertainties $\Delta \Phi$ because of two reasons: It is still a
challenging task to make a precise prediction of the objects' redshift. Thus,
the RLFs include high uncertainties for very distant objects. Nevertheless,
the functions
 are quite well known up to a redshift of $z\approx 2$. Since objects at a
higher redshift do not contribute significantly to the diffuse neutrino
spectrum, the error due to uncertainties in distance measurements will be
neglected compared to uncertainties in the jet-disk symbiosis model.

The jet-disk symbiosis model agrees with the data when allowing a scattering
of a factor of three in a $L_{radio}(L_{disk})$ graph. This transfers to an
uncertainty of $\Delta \phi_{\nu}=0.78\cdot\phi_{\nu}$, that means the
an uncertainty of a factor $\sim 2$ towards higher and a factor $\sim 5$ towards lower fluxes. 
The spectrum for flat spectrum sources includes an uncertainty of $\Delta
\phi_{\nu}^{f}=0.9\cdot \phi_{\nu}$ which means a
factor of $\sim 2$ above an a
factor of $10$ below the prediction.

Uncertainties due to the chosen cosmology would result in another factor of
two in the change of the normalization.

Since the maximum energy of the spectrum depends on the disk luminosity which
is converted into a radio luminosity using the jet-disk symbiosis model, the
cutoff may change by an order of magnitude at most.

\subsection{Discussion of a source flux from the supergalactic plane}
It has been stated before that the calculations include sources at $z>0.03$ to exclude possible anisotropies from the 
supergalactic plane. In the following paragraph, a possible flux from the supergalactic plane will be discussed. Note that 
this is just to make an estimate of a possible contribution from nearby sources and that final results do not include 
sources at $z<0.03$.
Assuming that the contribution to the total flux at $z<0.03$ is concentrated in the supergalactic plane yields an increase of
the flux by a factor 10 since the supergalactic plane,  [\cite{sg_plane}], covers approximately 10~\% of the sky. The ratio of 
the diffuse steep spectrum flux at $z>0.03$ and the flux from the supergalactic plane is of the order
\[
\frac{\Phi^{steep}(z<0.03)}{\Phi^{steep}(z>0.03)} \approx 0.01\,.
\]
Since this is only some percent of the total diffuse flux, it is negligible.
For flat spectrum sources, the result is 
\[
\frac{\Phi^{flat}(z<0.03)}{\Phi^{flat}(z>0.03)} \approx 0.02 
\]
and thus also negligible compared to the flux from outside the
supergalactic plane.
\subsection{Discussion of the final spectra \label{final_spectra}}
The comparison of the two
source type spectra considering only one neutrino flavor\footnote{The
  total flux is divided by a factor of three, taking into account
  neutrino oscillations.} in Fig.~\ref{flux_2_paper} shows that the dominant
contribution comes from steep spectrum sources assuming a particle
index of $p=2.0$. However, using $p=2.6$ for the steep source spectrum
reduces the contribution from this sample significantly as is
presented in Fig.~\ref{flux_26}. 
The cutoff in the energy spectrum at $E_{\nu}\sim 10^{9}$~GeV occurs since the 
maximum energy for each single source depends on the source luminosity as shown by~\cite{lovelace} and \cite{falcke1}. Thus, the
cutoff energy of the isotropic distribution depends on the sources' luminosity
distribution. The atmospheric neutrino spectrum
(dashed lines) dominates the spectrum up to
energies of $\sim 10^7$~GeV. 

The maximum optical depth of the sources can be estimated by assuming that the contribution
of the calculated spectrum has to be smaller than the limit set by AMANDA mentioned above. Thus, in case of an $E^{-2}$ spectrum
for steep spectrum sources, the optical depth can be restricted to
$\tau_{eff}<2$. This factor will grow significantly when using an $E^{-2.6}$
spectrum, i.e.~$\tau_{eff}<1000$. For flat spectrum sources, an effective
opacity of $\tau_{eff}<400$ can be estimated.

\begin{figure}
\begin{center}
\includegraphics[width=\linewidth]{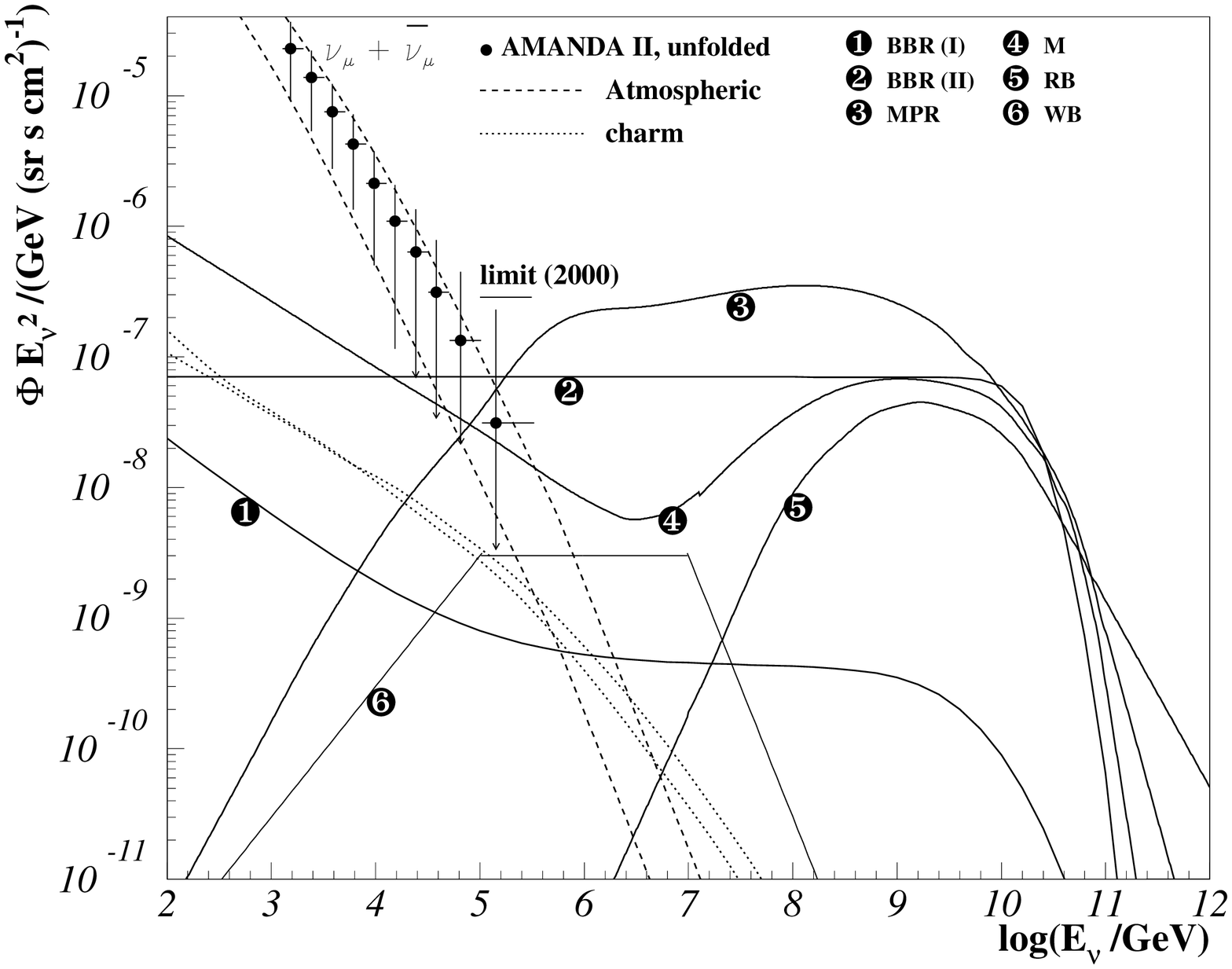}
\end{center}
\vspace{-0.5cm}
\caption{\textbf{$\nu_{\mu}+\overline{\nu}_{\mu}$ neutrino flux predictions.} (1) and (2) are the predictions done in this paper, BBR(I) using $p=2.6$ for 
the steep spectrum sources, BBR(II) applying $p=2$; (3):  $p\,\gamma$ interactions, MPR~[\cite{MPR}]; 
(4): M~[\cite{mannheimjet}], 
$p\,\gamma$ in blazar and $p\,p$ in its host galaxy; 
(5): RB~[\cite{rachen}], $p\,\gamma$;  
(6): WB~[\cite{WB}], diffuse neutrino flux from GRBs, assuming that GRBs are the sources of cosmic rays at $E>10^{19}$~eV. The dashed lines indicate the atmospheric neutrino spectrum
according to~\cite{volkova}, 
the upper line representing the horizontal flux, the lower line showing the vertical contribution. 
The dotted lines represent
a model for the atmospheric charm contribution as predicted in~[\cite{stasto}]. The uncertainties in the model occur due 
to larger errors in measurements of the Bj\"orken variable in forward scattering. Data points are AMANDA data showing the unfolded spectrum of 2000 data. It follows the atmospheric prediction. The limit on the extraterrestrial neutrino flux is set based on the data from the last bin of the unfolded spectrum 
[\cite{AMANDA}].}
\label{fluesse}
\end{figure}

A comparison of the results using two different spectral indices for the steep
spectrum sources makes clear how much the total flux contribution depends on the
spectral index of the sources. Changing the spectral index from $p=2.0$ to $p=2.6$ reduces the contribution from the source sample
by two orders of magnitude. In the pessimistic case, the detection of a signal from these types of AGN will require an array like
IceCube or Auger which are sensitive to UHE neutrinos at $\sim 10^{19}$~eV. In the case of an $E^{-2}$ spectrum the detection of
a diffuse signal should be possible in the near future. Fig.~\ref{fluesse} shows a summary of various extragalactic neutrino
flux predictions, see~[\cite{learned}] for a review. In the usual $E^{-2}$ scenario, the AGN spectrum should be detectable in the 
very near future, presumably already by running neutrino experiments like AMANDA. If the spectrum is steeper, however, it has 
been shown that a detection of a significant neutrino signal from AGN would need some years of operation of a km$^3$ experiment
like IceCube.
\clearpage
\section{Acknowledgments}
The authors would like to thank K.~Mannheim and H.~Falcke for valuable
discussions which led to much progress in realizing this work.

The work with P.~L.~Biermann has been supported through the AUGER theory and membership grant 05 CU1ERA/3 via DESY/BMBF 
(Germany). Further support for the work with P.~L.~Biermann has come from the DFG, DAAD, Humboldt Foundation (all Germany), 
grant 2000/06695-0 from FAPESP (Brasil) with G.~Medina-Tanco, a grant from KOSEF (Korea) through H.~Kang and D.~Ryu, a 
grant from ARC (Australia) through R.~J.~Protheroe, and a European INTAS grant with V.~Berezinsky. 
Special support comes from the European Sokrates/Erasmus grants in collaboration with East-European Universities, 
with partners M.~Ostrowski, K.~Petro\-vay, L.~Gergely, A.~Petrusel, and M.V.~Rusu, and VIHKOS through the FZ Karls\-ruhe. 
Current support comes from NATO for a collaboration with S.~Moiseenko and G.~Bisnovatyi-Kogan (Moscow), 
and the Chinese Academy of Sciences of China in work with Y.~Wang (Beijing/Nanjing). 

\clearpage
\small
\bibliography{literature}
\bibliographystyle{elsart-harv}
\end{document}